\documentclass[11pt, preprintnumbers, amsmath,amssymbre]{revtex4}
\usepackage{amsmath}
\usepackage{amsfonts}
\usepackage{amssymb}
\usepackage[colorlinks,linkcolor=blue,citecolor=red]{hyperref}
\usepackage{graphicx}
\usepackage{hyperref}
\usepackage{subfigure}
\usepackage{graphicx}% Include figure files
\usepackage{dcolumn}% Align table columns on decimal point
\usepackage{bm}% bold math
\usepackage{amsmath,amsthm,amsfonts}
\usepackage{graphicx}
\usepackage{physics}
\usepackage{hyperref}
\usepackage{lscape} %landscape pages support
\usepackage{mathtools}
\usepackage{appendix}
\graphicspath{{figs/}}

\newcommand{\nc}{\newcommand}
\nc{\ba}{\begin{eqnarray}}
\nc{\ea}{\end{eqnarray}}
\nc{\bfk}{\bf{k} }
\nc{\bfq}{\bf{q} }

\begin{document}

\title{ Vacuum Zero Point Energy and its Statistical Correlations in dS Background }

%\author{Authors$^{a}$}

\author{Hassan Firouzjahi$^{a}$}
\email{firouz@ipm.ir}

\author{Haidar Sheikhahmadi$^{a}$}
 \email{h.sh.ahmadi@gmail.com;h.sheikhahmadi@ipm.ir}

\affiliation{$^{a}$ School of Astronomy, Institute for Research in Fundamental Sciences (IPM),  P. O. Box 19395-5531, Tehran, Iran}

\begin{abstract}

We study the vacuum zero point energy associated to a scalar field with an arbitrary mass and conformal coupling in a dS background. Employing dimensional regularization scheme, we calculate the regularized  zero point energy density, pressure and the trace of the energy momentum tensor. It is shown that the classical relation  $\langle T \rangle =-4 \langle \rho \rangle$  for the vacuum stress energy tensor receives anomalous quantum correction  which depends on the mass and the conformal coupling while the relation $\langle \rho \rangle = - \langle P \rangle$ does hold.  We  calculate the density  contrast associated to the 
vacuum zero point energy  and show that  $\delta \rho \sim \langle \rho \rangle$ indicating an inhomogeneous and non-perturbative distribution of the zero point energy. Finally, we  calculate the skewness associated to the distribution of the  zero point energy and pressure and show that they are highly non-Gaussian.
\\

\end{abstract}

\maketitle

%\tableofcontents

\section{introduction}\label{Intro00}

Quantum field theory in a curved spacetime is a rich and vastly studied topic
which deals with important theoretical and observational phenomena \cite{DeWitt:1975ys, Birrell:1982ix, Fulling:1989nb, Mukhanov:2007zz, Parker:2009uva}. Since a concrete theory of quantum gravity is not at hand, usually one assumes that the background geometry is governed by the classical general relativity and then  the quantum fields are quantized in this classical background. This is a simplified and incomplete treatment of the full picture because the quantization of the gravitational degrees of freedom is expected  to be essential at very high energy (very short scales). But even within this simplified picture interesting and non-trivial effects emerge if one does not get too close to the quantum gravity scale.
For example, as the background may be dynamical, particle creation is a common phenomena in quantum field theory in a curved spacetime  \cite{Parker:1968mv, Parker:1969au,Parker:1971pt, Wald:1975kc, Panangaden:1977pc, Davies:1976ei,  Wilson:2019ago}. In addition, the concept of vacua is a non-trivial issue as different observers may define different vacua associated to their quantum fields
\cite{Hawking:1975vcx, Unruh:1976db, Unruh:1983ms, Jacobson:2003vx, Cozzella:2020gci, Firouzjahi:2022rtn}.

An important issue in studying quantum field theory in curved spacetime is the questions of regularization and renormalization. Similar to  quantum field theories in flat spacetime, physical quantities such as the energy momentum tensor, energy density and pressure suffer from infinities in a curved spacetime as well.  The fact that there is no unique vacuum in a curved spacetime while particles can be created add more complexities for the treatment of regularization and renormalization in a curved spacetime.
Therefore, it is an important question as how one can regularize the infinities and to read  off the finite physical quantities. There are various well established schemes for regularization and renormalization in curved spacetimes such as the point splitting regularization method  \cite{Christensen:1976vb, Christensen:1977jc, Davies:1977ze,Christensen:1978yd, Anderson:1990jh}, the adiabatic regularization method based on the WKB approximation \cite{ Parker:1974qw,Fulling:1974pu,Ford:1977in, Anderson:1987yt}, the zeta function regularization scheme \cite{Hawking:1976ja,Chadha:1977yd,Perry:1978jj,Elizalde:2002dd,Elizalde:1997hx} and the dimensional regularization procedure \cite{tHooft:1972tcz, tHooft:1974toh,
Bollini:1972ui,  Deser:1974cz, Dowker:1975tf, Barvinsky:1985an, Onemli:2002hr, Brunier:2004sb, Miao:2005am, Prokopec:2008gw, Miao:2010vs, Glavan:2021adm, Glavan:2020gal,
Zhang:2019urk,Ye:2022tgs}.
%and some other  techniques, which maybe are constructed by combination of some of these mentioned methods, see for example \cite{Zhang:2019urk,Ye:2022tgs}.

Understanding quantum field theory  in dS background is an important question both theoretically and observationally  \cite{Bunch:1978yq, Bunch:1978yw, Avis:1977yn, Anderson:1985cw, Armendariz-Picon:2023gyl}.
On the observational side, there are compelling evidences that the early universe experienced a period of inflation in which the background was nearly a dS background. The simplest models of inflation are based on scalar field dynamics in which a light scalar field rolls slowly on top of its nearly flat potential \cite{Weinberg:2008zzc, Baumann:2022mni}. While the background expansion is given by the potential, but there are quantum fluctuations associated to inflaton field
which are stretched to superhorizon scales. It is believed that these quantum perturbations are the seeds of large scale structures in universe and perturbations in CMB \cite{Kodama:1984ziu, Mukhanov:1990me}
which are well supported by cosmological observations \cite{Planck:2018vyg, Planck:2018jri}. In addition, various cosmological observations indicate that the universe is undergoing a phase of accelerated expansion now. The origin of dark energy as the source of the late time
acceleration is not known
but a cosmological constant associated with the vacuum zero point energy of fields is a prime candidate which fits the data well  \cite{Weinberg:1988cp, Sahni:1999gb, Peebles:2002gy, Copeland:2006wr, Martin:2012bt}. Beside being a possible candidate for the origin of dark energy, the vacuum zero point energy and its regularization in a dS spacetime 
is an important question of its own right \cite{ Onemli:2002hr, Brunier:2004sb, Miao:2005am, Prokopec:2008gw, Miao:2010vs, Glavan:2021adm, Glavan:2020gal}.

In this work we study the quantum fluctuations of a real scalar field with non-minimal coupling to gravity in a dS background, focusing on vacuum zero point energy and its statistical fluctuations.  While quantum field theory in a dS background has been studied extensively in the past but here we look at this question in a different perspective. Motivated by ideas from inflationary model building, here we pay particular attention on the statistical variations of the vacuum zero point energy density and pressure and look for their physical implications.

The rest of the paper is organized as follows. In section \ref{Mathematics} we present our setup while in section  \ref{dim-reg} we calculate the expectation values of the zero point energy and pressure. In section \ref{perturbations} we study the statistical fluctuations in energy density  and pressure and calculate their two-point and three-point correlation functions followed by Summary and Discussions in section \ref{summary}. Some technicalities dealing with higher correlations of the energy density and pressure  are presented in Appendix \ref{contractions}.

%%%%%%%%%%%%%%%%%%%%%%%%%%%%%%%%%%%%%%
\section{The Setup}
\label{Mathematics}

We consider a real scalar field $\Phi$ in a dS spacetime which is non-minimally coupled to gravity with the conformal coupling $\xi$. The action is given by
\begin{equation}
\label{Action00}
S=\int d^D x \sqrt{-g_{_D}}\left({-\frac{1}{2}\xi\Phi^2R} -\frac{1}{2} \nabla^\mu \Phi \nabla_ \mu \Phi-\frac{1}{2} m^2 \Phi^2\right)\,,
\end{equation}
in which $D$ refers to the dimension of the spacetime, $g_{_D}$ stands for the determinant of the metric and $m$ is the mass of the scalar field. Since we employ dimensional regularization to handle the quantum infinities, we keep the spacetime dimension general and only  at the end set $D=4-\epsilon$ with $\epsilon\rightarrow 0$ as in conventional dimensional regularization approach. In four dimensional spacetime  the theory is classically conformally invariant if $m=0$ and $\xi =\frac{1}{6}$. However, as it is well-known,  this classical symmetry is anomalous under quantum perturbations which will be studied in some details below.

We work in the test field limit where the background geometry is governed by the Einstein field equation and it is not affected by the presence of the test field. In order for this approximation to be consistent, we require the vacuum zero point energy and pressure associated to the fluctuations of $\Phi$ to be much smaller than the corresponding
background quantities.  These requirements put constraints on the mass of the test field as we shall study below.
To simplify the analysis, we consider a free theory in which there is no interaction in the field sector. It is an interesting question to extend the current analysis to the more physical setup where there is a self-interaction 
like $\lambda \Phi^4$ in the model. For works studying various aspects of quantum effects in models with $\lambda \Phi^4$ self-interaction see
\cite{Onemli:2002hr, Brunier:2004sb, Onemli:2004mb}.

The background geometry has the form of the FLRW metric,
\begin{equation}\label{metric}
d s^2=-d t^2+a(t)^2 d \mathbf{x}^2\,,
\end{equation}
where  $a(t)$ is the  scale factor and $t$ is the cosmic time. It is more convenient to work with the conformal time $\tau$ related to the cosmic time via $d \tau = dt/a(t)$
in terms of which the metric becomes conformally flat,
\ba
ds^2 = a(\tau)^2 \big( -d \tau^2 + d {\bf x}^2 \big) \, .
\ea
In terms of conformal time, the relation $a H \tau =-1 $ holds in a dS background which will be used frequently in the following analysis.

The dS spacetime is maximally symmetric, so the Ricci tensor and Ricci scalar are given as follows,
\ba
\label{maximal}
R_{\mu \nu} = (D-1)H^2 g_{\mu \nu} , \quad \quad R= D (D-1) H^2 \, ,
\ea
in which $H\equiv \frac{\dot a}{a} $ is the Hubble expansion rate during inflation.

The Klein-Gordon equation governing the dynamics of the field  is given by
\begin{equation}
\label{FieldEQ}
\Box\Phi  - \xi R\Phi  - m^2{\Phi ^2} = 0\,.
\end{equation}
To study  the quantum perturbations of the field, we  introduce the canonically normalized  field $\sigma(\tau)$
\ba
\mathrm{\sigma(\tau)}\equiv a^{\frac{D-2}{2}}\Phi(\tau) \, ,
\ea
in terms of which the action takes the following form,
\ba
\label{Action01}
S=\frac{1}{2} \int d \tau d^{D-1} {\bf x}\left[\sigma^{\prime}(\tau)^{2}-(\nabla\sigma)^2+\left(\frac{(D-4)(D-2)}{4}\big(\frac{a^{\prime}}{a}\big)^2+\frac{D-2}{2} \frac{a^{\prime \prime}}{a}-(m^2 +{\xi R}) a^2\right)\sigma^2\right] ,
\ea
where a prime indicates the derivative with respect to the conformal time.

%%%%%%%%%%%%%%%%%%%%%%%%%%%%%%%%%%%%%%%%%%
%\subsection{Mode Function }

To quantize the field, as usual,  we expand it in terms of the creation and annihilation operators  in the Fourier space as follows,
\begin{equation}
\label{quantized-sigma}
\sigma\left(x^\mu\right)=\int \frac{d^{D-1} \mathbf{k}}{(2 \pi)^{\frac{(D-1)}{ 2}}}\left(\sigma_k(\tau) e^{i \mathbf{k}\cdot\mathbf{x}} a_{\mathbf{k}}+\sigma^*_k(\tau) e^{-i \mathbf{k} \cdot \mathbf{x}} a_{\mathbf{k}}^{\dagger}\right)\,,
\end{equation}
where $\sigma_k(\tau)$ is the quantum mode  function while
$a_{\mathbf{k}}$ and $a_{\mathbf{k}}^{\dagger}$ satisfy the following
commutation relation in $D-1$ spatial dimension,
\begin{equation}
\label{Noncommutative}
\left[a_{\mathbf{k}}, a_{\mathbf{k^\prime}}^{\dagger}\right]=\delta^{D-1}(\mathbf{k}-\mathbf{k^\prime})\,.
\end{equation}

The equation of motion of the mode function  from the action \eqref{Action01} is given by
\begin{equation}
\label{EquationofMotion}
\sigma_k^{\prime \prime}(\tau)+\left[k^2+ \frac{1}{\tau^2} \Big(
\frac{m^2}{H^2 } + {{D (D-1)}} \xi-\frac{D(D-2)}{4} \Big) \right] \sigma_k(\tau)=0\,.
\end{equation}
Note that the above equation is similar to the Mukhanov-Sasaki equation associated to the  inflaton perturbations in an inflationary background. If we set $m=0$ and
$\xi =\frac{1}{6}$ in $D=4$, then the second term in the big bracket vanishes and the mode function reduces to its simple flat form. In a general $D$-dimensional spacetime with $m=0$, the conformal limit is attained for the special value of $\xi=\xi_D\equiv  \frac{D-2}{4 (D-1)} $. %given by $\xi_D\equiv \frac{D-2}{4 (D-1)}$.
%\ba
%\xi_D\equiv \frac{D-2}{4 (D-1)} \, .
%\ea 

Imposing the Bunch-Davies (Minkowski) vacuum deep inside the horizon, the
solution of the mode function from Eq. (\ref{EquationofMotion})
is given in terms of the Hankel function
\begin{equation}\label{Phi-k}
\Phi _k(\tau) = a^{\frac{{2 - D}}{2}}{\sigma _k}(\tau ) = {( - H\tau )^{\frac{{D - 1}}{2}}}{\left( {\frac{\pi }{{4H}}} \right)^{\frac{1}{2}}} {{e^{\frac{i \pi}{2}  (\nu + \frac{1}{2})} }}
H_\nu ^{(1)}( - k\tau ){\mkern 1mu}\,,
\end{equation}
where
\begin{equation}
\label{nu-00}
{{
\nu  \equiv  \frac{1}{2}{\mkern 1mu} \sqrt {(D-1)^2 {-4 D (D-1) \xi}- 4 \beta^2}\, , \quad \quad
\beta \equiv \frac{m}{H} }}\, .
\end{equation}
From the above expression of $\nu$ we see that it can be either real or pure imaginary depending on the values of $\xi$ and $\beta$. For a light field with $\beta <1$ and with a  moderate  value of $\xi$, the index  $\nu$ is real while for a heavy field with $\beta \gg 1$ we typically have a complex value of $\nu$. Both cases of real and imaginary $\nu$ will be considered in the following analysis.

%%%%%%%%%%%%%%%%%%%%%%%%%%%%%%%%%%%%%%%%%%
%\subsection{Energy momentum }

In our analysis below, we are mainly interested in the expectation values of the vacuum energy momentum tensor in vacuum $\langle T^{\mu \nu}_v \rangle$,
the vacuum zero point energy density $\langle \rho_v \rangle$, the vacuum zero point pressure
$\langle P_v \rangle$ and their statistical correlations.  In order to simplify the notation, we discard the subscript $v$ in the rest of analysis unless mentioned specifically.

The energy momentum tensor is given by,
\begin{eqnarray}\label{E-Mtensor-nonmimimal}\nonumber
{{T_{\mu \nu }}}& =& (1 - 2\xi ){\partial _\mu }\Phi {\partial _\nu }\Phi  + (2\xi  - \frac{1}{2}){g_{\mu \nu }}{g^{\alpha \beta }}{\partial _\alpha }\Phi {\partial _\beta }\Phi \\
\,\,\,\,\,\,\,\, &+ &\xi ({R_{\mu \nu }} - \frac{1}{2}{g_{\mu \nu }}R){\Phi ^2} + 2\xi ({g_{\mu \nu }}\Phi\Box \Phi  - \Phi {\nabla _\nu }{\nabla _\mu }\Phi ) - \frac{1}{2}{g_{\mu \nu }}m_\Phi ^2{\Phi ^2}\, .
\end{eqnarray}
Using the field equation (\ref{FieldEQ}) to eliminate $\Box \Phi$ combined with Eq. (\ref{maximal}),  $T_{\mu \nu}$ is simplified as follows
\ba
\label{T-simple}
{{T_{\mu \nu} = {\partial _\mu }\Phi {\partial _\nu }\Phi + \frac{g_{\mu \nu}}{2} ( 4 \xi -1)  \big( \partial^\alpha \Phi \partial_\alpha \Phi +  m^2
\Phi^2 \big) %\nonumber\\
+ \frac{\xi}{2} (D-1) \big( 2+ ( 4 \xi -1) D  \big) H^2 g_{\mu \nu} \Phi^2 - \xi
\nabla_\mu \nabla_\nu \Phi^2}}.  \nonumber
\ea
Similarly, the trace of the energy momentum-tensor  $T\equiv T^\mu_\mu$ is  given by
\ba
\label{T-eq}
T= 2 \Big( (D-1) \xi + \frac{2-D}{4} \Big) \Big( \partial^\alpha \Phi \partial_\alpha \Phi
+ D (D-1) \xi H^2 \Phi^2  \Big) + \Big( 2 \xi (D-1) -\frac{D}{2} \Big) m^2 \Phi^2 \, .
\ea

As we shall show explicitly below, $\langle \Phi^2 \rangle$ is independent of
$x^\mu$ so the vacuum expectation value of $\langle T_{\mu \nu} \rangle$ simplifies to
\begin{align}
\label{Tmunu-av}
{{ \langle T_{\mu \nu} \rangle  = \langle  {\partial _\mu }\Phi {\partial _\nu }\Phi  \rangle + \frac{g_{\mu \nu}}{2} ( 4 \xi -1)   \langle \partial^\alpha \Phi \partial_\alpha \Phi \rangle
+ \frac{g_{\mu \nu}}{2} \Big[ ( 4 \xi -1)  m^2
+ \frac{\xi}{2} (D-1) \big( 2+ ( 4 \xi -1) D  \big) H^2   \Big] \langle\Phi^2\rangle   }} .
 \nonumber
\end{align}
The vacuum zero point energy is $\rho = T_{00}$ so from the above expression we obtain
\ba
\label{rho-eq0}
\langle \rho \rangle  = \frac{(1+ 4 \xi)}{2} \langle \dot \Phi^2 \rangle +
\frac{(1- 4 \xi)}{2}  \langle   \nabla^i \Phi \nabla_i \Phi \rangle +
\frac{H^2}{2} \Big[  (1- 4 \xi)  {{\big( \beta^2 + D (D-1) \xi)}} - 2 (D-1) \xi  \Big] \langle \Phi^2 \rangle .
\ea
On the other hand,  the pressure $P$ is given by
\ba
P = \frac{1}{{D - 1}}{ \bot ^{\mu \nu }}{T_{\mu \nu }} \, ,
\ea
in which $\perp^{\mu \nu} \equiv g^{\mu \nu}+u^\mu u^\nu$ is projection operator and $u^\mu=(1,0,0,0)$ is the comoving velocity. Consequently, we obtain
\ba
\label{P-eq}
P= \frac{1}{ {D - 1} }( T+ \rho) \, .
\ea

%%%%%%%%%%%%%%%%%%%%%%%%%%%%%%%%%%%%%%%%%%
\section{Dimensional Regularizations and Expectation Values }
\label{dim-reg}

In this section we calculate $\langle \rho \rangle$ and $\langle P \rangle$ using dimensional regularization scheme in $D$ dimension.

From Eq. (\ref{rho-eq0}) we see that $\langle \rho \rangle$ contains the following three ingredients:
\ba
\label{rhoi}
\rho_1 \equiv \frac{1}{2} \dot \Phi^2 \, , \quad \quad
\rho_2 \equiv \frac{1}{2} g^{i j} \nabla_i \Phi \nabla_j \Phi \, , \quad \quad
\rho_3 \equiv \frac{1}{2} H^2 \Phi^2 \, .
\ea

Let us start with $\langle \rho_1\rangle $. With the mode function given in Eq. (\ref{quantized-sigma})
and performing a simple  contraction using the commutation relation (\ref{Noncommutative}) we obtain
\ba
\langle \rho_1\rangle  =\frac{ \mu ^{4 - D} }{2a^2(\tau ) }
\int \frac{ d^{D - 1} {\bf k}}{(2\pi )^{D - 1} }  \left| \Phi _k^\prime (\tau ) \right|^2 \, ,
\ea
in which $\mu$ is a mass scale to keep track of the dimensionality of physical quantities  as usual in dimensional regularization analysis.

To proceed further, we decompose the integral into the radial and angular parts as follows
\ba
{{\rm{d}}^{D-1}}{\bfk} = {k^{D - 2}}\;{\rm{d}}k\; {{{\rm{d}^{D-2}} }}\Omega {\mkern 1mu}\, ,
\ea
in which  ${\rm{d}^{D-2}}\Omega$ represents the $D-2$-dimensional  angular part
with the volume
\ba
 \int \mathrm{d}^{D-2} \Omega=\frac{ 2 \,  \pi^{\frac{D-1}{2}}}{\Gamma\left(\frac{D-1}{2}\right)}\,.
\ea
Combining all numerical factors and defining the dimensionless variable $x\equiv - k \tau$ we finally obtain
\ba
\label{rho1}
\langle \rho_1 \rangle =
{\frac {{\pi}^{\frac{3-D}{2}  }{\mu}^{4-D}{H}^{D}}{{2}^{1+D}\Gamma \left( \frac{D-1}{2} \right) }} e^{-\pi \mathrm{Im}(\nu)}  \int_0^{\infty} dx~x  \left|\frac{d}{d x}\left(x^\frac{D-1}{2} H_{ \nu}^{(1)}(x)\right)\right|^2 \, ,
\ea
where after integrating it reads \footnote{We use the Maple computational software to calculate the integrals in Eqs. (\ref{rho1}), (\ref{rho2}) and (\ref{rho3}).}
\begin{eqnarray}\label{rho1-int}
\left\langle {{\rho _1}} \right\rangle  = \frac{{{\mu ^{4 - D}}{\pi ^{ - \frac{D}{2} - 1}}}}{4} {\mkern 1mu} \Gamma \Big( {\nu  + \frac{D}{2} + \frac{1}{2}} \Big)\Gamma \Big( { - \nu  + \frac{D}{2} + \frac{1}{2}} \Big)\Gamma \big( { - \frac{D}{2}} \big) \cos \big( {\pi {\mkern 1mu} \nu } \big)   {\big( {\frac{H}{2}} \big)^D}\,.
\end{eqnarray}
As explained previously, $\nu$ can be either real or pure imaginary. In the latter case, one simply replaces $\nu$ by $i \nu$ in the above and in the following expressions. In addition, the following relation for the complex conjugation  $\overline{H_{ i\nu}^{(1)}}(x)$
holds
\ba
\overline{H_{ i\nu}^{(1)}}(x) = e^{\pi \mathrm{Im}(\nu)}  H_{ i \nu}^{(2)}(x) \, ,
\ea
which was used to obtain Eq. (\ref{rho1-int}).

Following similar steps, for the remaining components we obtain
\ba
\label{rho2}
\langle \rho_2 \rangle =
{\frac {{\pi}^{\frac{3-D}{2}  }{\mu}^{4-D}{H}^{D}}{{2}^{1+D}\Gamma \left( \frac{D-1}{2} \right) }} e^{-\pi \mathrm{Im}(\nu)}  \int_0^{\infty} dx~x^D  \left| H_{ \nu}^{(1)}(x)\right|^2 \, ,
\ea
in which the result can be expressed as
\begin{eqnarray}\label{rho2-int}
\left\langle {{\rho _2}} \right\rangle  = \frac{{{\mu ^{4 - D}}{\pi ^{ - \frac{D}{2} - 1}}}}{4} {\mkern 1mu} \big( {D - 1 } \big) \Gamma \Big( {\nu  + \frac{D}{2} + \frac{1}{2}} \Big)\Gamma \Big( { - \nu  + \frac{D}{2} + \frac{1}{2}} \Big)\Gamma \big( { - \frac{D}{2}} \big) \cos \big( {\pi {\mkern 1mu} \nu } \big)   {\big( {\frac{H}{2}} \big)^D}\,,
\end{eqnarray}
and
\ba
\label{rho3}
\langle \rho_3 \rangle =
{\frac {{\pi}^{\frac{3-D}{2}  }{\mu}^{4-D}{H}^{D}}{{2}^{1+D}\Gamma \left( \frac{D-1}{2} \right) }} e^{-\pi \mathrm{Im}(\nu)}  \int_0^{\infty} dx~x^{D-2}  \left| H_{ \nu}^{(1)}(x)\right|^2 \, ,
\ea
that yields
\begin{eqnarray}\label{rho3-int}
\left\langle {{\rho _3}} \right\rangle  = \frac{{{\mu ^{4 - D}}{\pi ^{ - \frac{D}{2} - 1}}}}{2} \Gamma \Big( {\nu  + \frac{D}{2} - \frac{1}{2}} \Big)\Gamma \Big( { - \nu  + \frac{D}{2} - \frac{1}{2}} \Big)\Gamma \Big( { - \frac{D}{2} + 1} \Big)\cos \big( {\pi {\mkern 1mu} \nu } \big){\big( {\frac{H}{2}} \big)^D}\,.
\end{eqnarray}
Incidentally, from the above equations we see that $\langle \rho_i \rangle $ are constants so  $\langle \Phi^2(x) \rangle \propto \langle \rho_3 \rangle $ is constant as advertised previously. This is consistent with the fact that the dS background is a maximally symmetric spacetime so $\langle \Phi^2(x)\rangle $ is expected  to be a constant. 

With the component of $\langle \rho_i \rangle$ given in Eqs. (\ref{rho1-int}), (\ref{rho2-int}) and (\ref{rho3-int}) in a general $D$-dimensional dS spacetime we obtain the following relations among them,
\ba
\label{rho13}
\langle \rho_1 \rangle = \Big[ (D-1) \xi + \frac{\beta^2}{D} \Big] \langle \rho_3 \rangle \, , \quad \quad \langle \rho_2 \rangle = -(D-1) \langle \rho_1 \rangle =
-\Big[ (D-1)^2 \xi + \frac{(D-1)}{D}  \beta^2 \Big] \langle \rho_3 \rangle  \, .
\ea
The above relations between $\langle \rho_i \rangle$ will be very helpful in the following analysis.
%\ba \label{rho23} \langle \rho_2 \rangle = -\Big[ (D-1)^2 \xi + \frac{(D-1)}{D}  \beta^2 \Big] \langle \rho_3 \rangle  \, . \ea

Having each components of $\langle \rho_i \rangle$ at hand, the zero point energy $\langle \rho \rangle$ from Eq. (\ref{rho-eq0})  is given by
\ba
\label{rho-eq}
{{\langle \rho \rangle = (1+ 4 \xi) \langle \rho_1 \rangle +  (1- 4 \xi) \langle \rho_2
 \rangle + \Big[  (1- 4 \xi)  {{\big( \beta^2 + D (D-1) \xi)}} - 2 (D-1) \xi  \Big] \langle \rho_3 \rangle }}\, .
\ea
Using the relation given in Eq. (\ref{rho13}) this further simplifies to
\ba
\label{rho-3}
\langle \rho \rangle = \frac{2 \beta^2}{D}  \langle \rho_3 \rangle \, .
\ea

Similarly, the expectation value  $\langle T\rangle $ from Eq. (\ref{T-eq}) is obtained
to be
\ba
\label{T-eq2}
\langle T \rangle &=& \Big[ ( D-2) - 4 (D-1) \xi \Big] \Big( \langle \rho_1 \rangle  - \langle \rho_2 \rangle  - D (D-1)  \xi \langle \rho_3 \rangle  \Big) + \Big[ 4 (D-1)\xi  -  D  \Big] \beta^2  \langle \rho_3 \rangle \nonumber\\
&=&- 2 \beta^2 \langle \rho_3 \rangle \, .
\ea
Comparing to Eq. (\ref{rho-3}) we obtain the interesting result that
\ba
\label{T-val1}
\langle T \rangle= -D \langle \rho \rangle \, .
\ea

It is crucial to note that the above relations are valid for a general value of $D$. In particular, we should not set $D=4$ in the above relations before performing dimensional regularizations. The reason is that we work in $D=4- \epsilon$ dimension in which $\langle \rho_i \rangle$, $\langle \rho \rangle$ and 
$\langle T \rangle$ will contain the divergent $\frac{1}{\epsilon}$ terms plus the regular terms. Consequently, there will be additional finite contributions  from the products of a function of $D$ and each of
$\langle \rho_i \rangle$ or $\langle \rho \rangle$. Intuitively speaking, these are
``quantum anomalies" which can not be seen classically. More specifically, 
consider the relation between $\langle T \rangle$  and $\langle \rho \rangle$ in 
Eq. (\ref{T-val1}). If we simply set $D=4$, we obtain the classical result 
$\langle T \rangle= -4 \langle \rho \rangle$ which is expected for the vacuum zero point energy.  
However, a careful investigation shows that this is not true. Indeed, as we shall verify below, setting $D=4 - \epsilon$ and performing the dimensional regularization to leading order  we have
\ba
\label{T-val2}
\langle T \rangle+4 \langle \rho \rangle={\cal A} \, ,
\quad \quad  \quad {\cal A} \equiv -\frac{H^4 \beta^2}{32 \pi^2} ( \beta^2 + 12 \xi -2) \, .
 \ea
The quantity ${\cal A}$ is a common effect which signals the  quantum ``anomalous'' contributions. We see that in the massless limit $\beta=0$  the above anomalous contribution vanishes.  Furthermore, for a massive field  if $\beta$ and $\xi$ arrange such that $\xi =\frac{1}{6} -\frac{\beta^2}{12}$ the anomalous contribution vanishes as well.

Similarly, the relation between $\langle \rho \rangle$  and $\langle \rho_3 \rangle$ in Eq. (\ref{rho-3}) receives the anomalous correction, yielding 
%If we simply set $D=4$, we obtain $\langle \rho \rangle= \frac{\beta^2}{2} \langle \rho_3 \rangle$. However, a careful investigation shows that this is not true. Indeed, as we shall verify below, we have
\ba
\label{anomaly}
\langle \rho \rangle- \frac{\beta^2}{2} \langle \rho_3 \rangle= \frac{{\cal A}}{4}\, . %\quad \quad  \quad{\cal A} \equiv -\frac{H^4 \beta^2}{32 \pi^2} ( \beta^2 + 12 \xi -2) \, .
\ea

On the other hand, from Eq. (\ref{P-eq}), combined with Eq. (\ref{T-val1}), we obtain the following relation between $\langle P \rangle$ and $\langle \rho \rangle$:
 \ba
 \label{rho-P}
 \langle P \rangle = - \langle \rho \rangle \, .
 \ea
The above relation between $ \langle P \rangle$ and $ \langle \rho \rangle$
is exact and is anomalous free. It holds for both massive  and massless fields. Physically this makes sense since we are dealing with bubble diagrams. As the spacetime is locally Lorentz invariant, then one requires \cite{Weinberg:1988cp, Martin:2012bt, Firouzjahi:2022xxb}
$\langle T_{\mu   \nu} \rangle = -\langle \rho \rangle g_{\mu\nu}$ which also yields  $\langle P \rangle = - \langle \rho \rangle $. Now contracting this tensorial relation with $g^{\mu \nu}$ we obtain Eq. (\ref{T-val1}). However, as mentioned previously,  Eq. (\ref{T-val1}) does not mean that $\langle T \rangle=  -  \langle \rho \rangle + 3  \langle P \rangle= 
-4 \langle \rho \rangle$\ since $D=4-\epsilon$ and there are divergent $\frac{1}{\epsilon}$ terms hiding inside $\langle \rho \rangle$. Intuitively speaking, local Lorentz invariance in dimensional regularization scheme adds a new ``extra dimension" of size $\epsilon$ which causes the anomalous relation  Eq. (\ref{T-val2}).

Now, let us calculate $\langle \rho \rangle$ from Eq. (\ref{rho-3})
with the value of $\rho_3$ given in Eq. (\ref{rho3-int}). Performing the dimensional regularization  to relevant order we obtain
\ba
\label{rho-val}
\langle \rho \rangle = {\cal A} \big( \frac{-1}{\epsilon}+ \frac{\Delta }{2} \big)
+\frac{H^4 \beta^2}{128 \pi^2} (2 - 8 \xi - 3 \beta^2) \, ,
\ea
in which $\Delta$ is another common factor defined via
\ba
\label{Delta}
\Delta \equiv \ln \Big( \frac{H^2}{4\pi \mu^2 }  \Big) + 2\Psi(\nu+ \frac{1}{2})  -  \pi \tan( \nu \pi) \, ,
\ea
where  $\Psi(x)$ is the digamma function and we have shifted $\mu$ by $\gamma$, the Euler number, which does not affect the physical result.
Furthermore, after performing the dimensional regularization analysis we can now
set  $D=4$ in which  from Eq. (\ref{nu-00}) we obtain
\ba
\label{nu-val}
\nu = \frac{1}{2}\sqrt{9- 4 \beta^2 - 48 \xi} \, .
\ea
In particular, for the special case of $\beta= \xi=0$, we obtain the expected result $\nu =\frac{3}{2}$ for a massless field in the dS background. In addition, for a heavy field with $\beta \gg 1$, $\nu$ becomes pure imaginary.

Looking at Eq. (\ref{rho-val}) we see that the parameter ${\cal A}$ is the coefficient of the divergent $\frac{1}{\epsilon}$ term so that is why we obtain the hidden anomalous contribution in Eqs. (\ref{T-val2}) and (\ref{anomaly}) when expanding
$D= 4- \epsilon$.

Finally, after subtracting the divergent $\frac{1}{\epsilon}$ term  in Eq. (\ref{rho-val}) via the appropriate counter terms, the regularized value of the zero point energy is obtained to be
\ba
\label{rho-reg}
\langle \rho \rangle_{\mathrm{reg}}  &=&  \frac{  {\cal A} \, \Delta }{2}
+\frac{H^4 \beta^2}{128 \pi^2}
(2 - 8 \xi - 3 \beta^2)  \\
&=& \frac{H^4 \beta^2}{64 \pi^2} \Big \{ ( \beta^2 + 12 \xi -2)
\Big[ \ln\Big( \frac{H^2}{4\pi \mu^2 }  \Big) + 2\Psi(\nu+ \frac{1}{2})  -  \pi \tan( \nu \pi)  \Big] + 1 - 4 \xi - \frac{3}{2} \beta^2  \Big \} \, . \nonumber
\ea

As is common in dimensional regularization approach the term  $\ln\Big( \frac{H}{ \mu }  \Big)$ originates from the  regularization. To read off the physical contribution, one has to further renormalize the above finite term. This can be achieved  upon choosing a physical value for the mass scale parameter $\mu$ or if one compares the values of
$\langle \rho \rangle_{\mathrm{reg}}$ at two different energy scales and look for its running with the change of the energy scale.

As explained previously, depending on the mass of the field, the index $\nu$ in Eq. (\ref{nu-val})
can be either real or imaginary. The former happens typically when the field is light or $\xi$ is not large while the latter corresponds to the case where the field is heavy with $\beta \gg 1$. Below we study each case separately.

%%%%%%%%%%%%%%%%%%%%%%%%%%%%%%%%%%%%%%%%%%
\subsection{Light field with real $\nu$}
\label{light}

Now we consider the case where the field is light enough so $\nu$ is real.
A particular case of interest is the massless limit $\beta=0$. We may also consider different limit of $\xi$ as well, such as the special cases $\xi=0$ and the conformal limit $\xi =\frac{1}{6}$.

From Eq. (\ref{rho-reg}) it may look that for massless field with $\beta=0$, we obtain $\langle \rho \rangle_{\mathrm{reg}}=0$. However, this is tricky as there is a particular limit in which the function $\tan (\nu \pi)$ diverges
when both $\beta, \xi \rightarrow 0$.  Taking the limit $\beta, \xi \rightarrow 0$ properly, we obtain
\ba
\label{massless-rho}
\langle \rho \rangle_{\mathrm{reg}}= \frac{3 H^4}{32 \pi^2} \, ,
 \quad \quad \quad  (\xi=\beta=0) \, .
\ea

Another limit of interest is  $\xi \ll 1$ such  that  $\xi \ll \beta^2 <1$. In this limit  we obtain
\ba
\langle \rho \rangle_{\mathrm{reg}}\simeq \frac{3 H^4}{32 \pi^2} - \frac{9 \xi H^4}{8 \pi^2\beta^2}
 -\frac{ H^4 \beta^2}{32 \pi^2} \Big[ \ln\Big( \frac{H^2}{4\pi \mu^2 }  \Big)  + \frac{10}{3} \Big]  +  \frac{ H^4 \beta^4}{64 \pi^2} \Big[ \ln\Big( \frac{H^2}{4\pi \mu^2 }  \Big)  - \frac{31}{54} \Big]  %+  {\cal O}( \beta^6)
\, , \quad
(\xi\ll \beta^2), \nonumber
\ea
in which the subleading terms of orders $\xi^2 \beta^{-4}$ or $\beta^6$ and higher orders are neglected in the above expansion.  On the other hand, for larger values  of $\xi$, we obtain $\langle \rho \rangle_{\mathrm{reg}} \propto \beta^2$ with the coefficient depending on the value of $\xi$. For example, for the particular limit with  $\xi = \frac{1}{6}$ we obtain
\ba
\langle \rho \rangle_{\mathrm{reg}}= -\frac{ H^4}{96 \pi^2} \beta^2
 + \frac{ H^4}{64 \pi^2} \Big[ \ln\Big( \frac{H^2}{4\pi \mu^2 }  \Big)  - \frac{1}{2} \Big] \beta^4 +  {\cal O}( \beta^6) \, , \quad \quad \quad
(\xi=\frac{1}{6} ) \, .
\ea
If we further assume that $\beta=0$ so the theory is classically conformal (with $m=0$ and $\xi = \frac{1}{6}$),
then the above expression yields $\langle \rho \rangle_{\mathrm{reg}} =0 $.

Similarly, for $ \langle T \rangle_{\mathrm{reg}}$ we can use the anomalous relation
(\ref{T-val2}) to obtain
\ba
 \langle T \rangle_{\mathrm{reg}} &=& -4  \langle \rho \rangle_{\mathrm{reg}}
 + {\cal A} \nonumber\\
 &=& ( 1- 2 \Delta )  {\cal A} - \frac{H^4 \beta^2}{32 \pi^2} (2 - 8 \xi - 3 \beta^2) \, .
\ea
For the particular case of $\xi=\beta=0$ we obtain
\ba
\langle T \rangle_{\mathrm{reg}}= -\frac{3 H^4}{8 \pi^2} \, ,
 \quad \quad \quad  (\xi=\beta=0) \, .
\ea
Curiously we see the trace anomaly in which $\langle T \rangle_{\mathrm{reg}} \propto H^4 \propto R^2 \neq 0$. This is the hallmark of quantum field theory in a curved spacetime \cite{Bunch:1978yq}.  For small value of $\xi$ with $\xi \ll \beta^2$, we obtain
\ba
\langle T \rangle_{\mathrm{reg}}\simeq \frac{-3 H^4}{8 \pi^2} + \frac{9 \xi H^4}{2 \pi^2\beta^2}  +\frac{ H^4 \beta^2}{8 \pi^2} \Big[ \ln\Big( \frac{H^2}{4\pi \mu^2 }  \Big)  + \frac{23}{6} \Big]  -  \frac{ H^4 \beta^4}{16 \pi^2} \Big[ \ln\Big( \frac{H^2}{4\pi \mu^2 }  \Big)  - \frac{2}{27} \Big]   \, , \quad   (\xi\ll \beta^2). \nonumber
\ea

On the other hand, for the particular case $\xi\rightarrow \frac{1}{6}$  we obtain
\ba
\langle T \rangle_{\mathrm{reg}} =   \frac{ H^4 \beta^2}{24 \pi^2}
 -  \frac{3 (\xi - \frac{1}{6})  }{4 \pi^2} H^4 \beta^2 \Big[ \ln\Big( \frac{H^2}{4\pi \mu^2 }  \Big)   + \frac{7}{8} \Big]  %+  {\cal O}( \beta^4)
 +  {\cal O}( \big(\xi - \frac{1}{6} \big)^2, \beta^4 )
 \, , \quad \quad
(\xi \rightarrow \frac{1}{6} ) \, .
\ea
If we further assume $\beta=0$ so the theory is classically conformal invariant
(with $m=0$ and $\xi =\frac{1}{6}$), then  $\langle T \rangle_{\mathrm{reg}} = 0$. This shows that there is no trace anomaly in the quantum level when the theory is classically conformal invariant.  This is in contrast with the result of \cite{Bunch:1978yq} who obtained $\langle T \rangle_{\mathrm{reg}} \propto R^2 \propto H^4$
when $\xi =\frac{1}{6}$ and $\beta=0$.

%%%%%%%%%%%%%%%%%%%%%%%%%%%%%%%%%%%%%%%%%%
\subsection{Heavy Field with Imaginary $\nu$}
\label{heavy}

For the heavy field with $\beta \gg 1$, the index $\nu$ in Eq. (\ref{nu-val})
becomes pure imaginary. All our results such as Eq. (\ref{rho-reg}) are formally valid with the understanding that  $\nu \equiv i \nu_0$ with
\ba
\label{nu0}
\nu_0 \equiv \frac{1}{2}\sqrt{4 \beta^2 + 48 \xi - 9} \, \simeq \beta \, .
\ea
Correspondingly, Eq. (\ref{nu-val}) yields
\ba
\label{rho-reg-heavy}
\langle \rho \rangle_{\mathrm{reg}}
= \frac{H^4 \beta^2}{64 \pi^2} \Big\{
( \beta^2 + 12 \xi -2)
\Big[ \ln\big( \frac{H^2}{4\pi \mu^2 }  \big) + 2\Psi(i\nu_0+ \frac{1}{2})  - i \pi \tanh( \nu_0 \pi) \Big] + 1 - 4 \xi - \frac{3  \beta^2}{2}  \Big\}
\ea
In the limit $\nu_0 \gg 1$, we have
\ba
2\Psi(i\nu_0+ \frac{1}{2})  - i \pi \tanh( \nu_0 \pi) \rightarrow 2 \ln(\nu_0)
+ {\cal O}(\nu_0^{-2} )\, .
\ea
Plugging this relation into Eq. (\ref{rho-reg-heavy}), assuming that $\beta \gg \xi$ and shifting the mass scale  $\mu $ by a constant value,
we obtain
\ba
\label{rho-reg-heavy2}
\langle \rho \rangle_{\mathrm{reg}}
= \frac{H^4 \beta^4}{64 \pi^2}  \ln\big( \frac{\nu_0^2 H^2}{4\pi \mu^2 }  \big)
+{ \cal O} ( \beta^2 H^4) \, .
%\equiv \frac{H^4 \beta^4}{64 \pi^2}  \ln\big( \frac{H^2}{4\pi \tilde \mu^2 }  \big)
\ea
Now noting that $\nu_0 \simeq \beta = \frac{m}{H} $, we obtain
\ba
\label{rho-reg-heavy2}
\langle \rho \rangle_{\mathrm{reg}}
= \frac{m^4}{64 \pi^2}  \ln\big( \frac{m^2}{4\pi \mu^2 }  \big)
+{ \cal O} ( m^2 H^2) \, .
\ea
The above result  agrees with the  vacuum zero point energy density in flat background  \cite{Martin:2012bt, Firouzjahi:2022xxb, Akhmedov:2002ts, Koksma:2011cq, Ossola:2003ku, Visser:2016mtr}.  This result is also obtained in the black hole background \cite{Firouzjahi:2022vij} when the Compton wavelength of the field is much smaller than the Schwarzschild radius of the black hole. As argued in \cite{Martin:2012bt} and \cite{Firouzjahi:2022xxb} one expects that $\langle \rho_v\rangle$ for a heavy field 
in a curved background agrees with the corresponding result in a flat background. The reason  is that the energy density is a local property of the spacetime. 
Since the Lorentz invariance is a local symmetry in GR, then the equivalence principle requires that $\langle \rho_v\rangle$ for a heavy field in a curved background,  with the Compton wavelength much smaller than the curvature radius of the spacetime,  agrees with $\langle \rho_v\rangle$ in a flat background. Nonetheless, it is an interesting exercise to demonstrate this physical expectation explicitly as we showed above.

Since we work in the test field limit, we have to make sure that the induced vacuum energy density  from quantum fluctuations does not affect the background geometry.
For this to be the case, we require $\langle \rho \rangle_{\mathrm{reg}}  \ll 3 M_P^2 H^2$ in which $M_P$ is the reduced Planck mass. Correspondingly,  this absence of the backreaction
imposes the following upper bound on the mass of the quantum field
\ba
\label{beta-bound}
\beta < \sqrt{\frac{M_P}{H} } \, .
\ea
This is an interesting bound. For example, suppose the background dS represents  an inflationary universe. This is a good approximation as during inflation the background is very nearly like a dS spacetime. Upper bound on the amplitude of primordial tensor perturbations from the Planck observation \cite{Planck:2018vyg, Planck:2018jri} requires that $H \lesssim 10^{-6} M_P$. This imposes the bound  $\beta < 10^{3}$ in order for our heavy field to remain a test field during inflation. Superheavy field with $\beta$ much larger than the bound given in Eq. (\ref{beta-bound}) would modify the background geometry and one has to solve the mode function with these corrections included.

%%%%%%%%%%%%%%%%%%%%%%%%%%%%%%%%%%%%%%%%%%
\section{Density Contrast and Skewness}
\label{perturbations}

In the previous analysis we have calculated the average physical quantities such as
$\langle \rho \rangle$. However, as the quantum field is fluctuating, there are fluctuations in $\rho $ as well.  In this section we calculate the variance in the energy density and pressure and their contrasts, i.e. $\frac{\delta \rho}{\langle \rho \rangle}$
and $\frac{\delta P}{\langle P \rangle}$.   As the analysis are complicated, we restrict ourselves to the spacial case $\xi=0$ but for arbitrary value of $\beta$. In addition, we also calculate the skewness which is a measure of the non-Gaussian distribution of the energy momentum tensor field.

To simplify the notation, let us absorb the parameter $\beta$ into $\rho_3$ by defining
\ba
\tilde \rho_3 \equiv \beta^2 \rho_3 = \frac{m^2}{2} \Phi^2 \, .
\ea
Then setting $\xi=0$ the energy density $\rho$ is simply given by
\ba
\rho= \rho_1 + \rho_2 + \tilde \rho_3 \, ,
\ea
while  Eqs. (\ref{rho13})  and (\ref{rho-3}) yield the the following relations among $\langle \rho_i\rangle$,
\ba
\label{relations}
\langle \rho \rangle = 2  \langle \rho_1 \rangle  = \frac{-2}{D-1} \langle \rho_2 \rangle
=  \frac{2}{D} \langle \tilde \rho_3 \rangle \, .
\ea
As explained previously, it is important that we do not set $D=4$ at this stage.
We set  $D=4$ only at the end of dimensional regularization where the divergent term and the leading finite terms are extracted from the analysis.

%%%%%%%%%%%%%%%%%%%%%%%%%%%%%%%%%%%%%%%%%%
\subsection{Density contrast }\label{Variancedelta2}

We are interested in  the variance $\delta \rho^2 \equiv \langle \rho^2 \rangle - \langle \rho \rangle^2$ which is given by
\begin{equation}\label{deltarhosquared}
\begin{aligned}
\delta \rho^2 & =\left\langle\rho_1^2\right\rangle+\left\langle\rho_2^2\right\rangle+\left\langle\rho_3^2\right\rangle+\left\langle\rho_1 \rho_2\right\rangle+\left\langle\rho_2 \rho_1\right\rangle+\left\langle\rho_1 \rho_3\right\rangle+\left\langle\rho_3 \rho_1\right\rangle+\left\langle\rho_2 \rho_3\right\rangle+\left\langle\rho_3 \rho_2\right\rangle \\
& -\left(\left\langle\rho_1\right\rangle^2+\left\langle\rho_2\right\rangle^2+\left\langle\rho_3\right\rangle^2\right)-2\left\langle\rho_1\right\rangle\left\langle\rho_2\right\rangle-2\left\langle\rho_1\right\rangle\left\langle\rho_3\right\rangle-2\left\langle\rho_2\right\rangle\left\langle\rho_3\right\rangle .
\end{aligned}
\end{equation}
To proceed further we need to calculate $\langle \rho_i \rho_j \rangle$.
Performing various contractions (see Appendix \ref{contractions} for further details),
one can show that
\ba\label{rho1andrho3}
\langle \rho_1^2 \rangle = 3  \langle \rho_1 \rangle^2 \, ,
\quad \quad
 \langle \tilde \rho_3^{\,2} \rangle = 3  \langle \tilde \rho_3 \rangle^2 \, .
\ea
The above results are understandable since $\Phi$ is a Gaussian field while
$\rho_1, \rho_3$ are made of the quartic powers of $\Phi$ so the Wick contractions yield the factor 3 above. On the other hand, the expectation value $\langle \rho_2^2 \rangle$ is somewhat non-trivial as we have the integration of the components of the momentum in $D-1$-dimensional space.
Performing the appropriate  contractions and the integrations
 over the momentum, we obtain (see Appendix \ref{contractions} for further details)
 \ba\label{rho2-variance}
 \langle \rho_2^2 \rangle = \big( 1+ \frac{2}{D-1} \big)   \langle \rho_2 \rangle^2 \, .
 \ea
%Note the non-trivial form of $\langle \rho_2^2 \rangle$ compared to other terms which arise from integrating the components of the momentum in $D$-dimensional spacetime.

 From the above relations for $\langle \rho_i^2 \rangle$ we obtain
 \ba
 \label{var-i}
 \delta \rho_1^2 = 2 \langle \rho_1 \rangle^2 \, , \quad \quad
 \delta \rho_2^2 = \frac{2}{D-1} \langle \rho_2 \rangle^2 \, , \quad \quad
 \delta \tilde\rho_3^2 = 2 \langle \tilde \rho_3 \rangle^2 \, , \quad \quad
 \ea
 which will be useful later on.

On the other hand, one can check that the average of the cross terms
$\langle \rho_i \rho_j\rangle$ with $i \neq j$ commute:
\ba
\label{rhotimessquared}
\langle \rho_1 \rho_2\rangle=\langle\rho_1 \rangle\langle\rho_2 \rangle\,,~~~~~\langle \rho_1 \tilde \rho_3\rangle=\langle\rho_1 \rangle\langle \tilde\rho_3\rangle\,,~~~~~~ \langle \rho_2  \tilde \rho_3\rangle=\langle\rho_2 \rangle\langle \tilde \rho_3 \rangle\,.
\ea
Correspondingly, the variance $\delta \rho^2$ is obtained to be the sum of the variances associated to  individual contributions:
\ba
\delta \rho^2 &=&  \delta \rho_1^2+ \delta \rho_2^2 + \delta \tilde\rho_3^2
\nonumber\\
&=& 2 \big( \langle \rho_1 \rangle^2 + \langle \tilde \rho_3 \rangle^2 \big)
+  \frac{2}{D-1}  \langle \rho_2 \rangle^2 \nonumber\\
&=&  \Big[ 2\big( \frac{1}{4} + \frac{D^2}{4} \big) +  \frac{2}{D-1}  \frac{(D-1)^2}{4}
\Big] \langle \rho \rangle^2 \, ,
\ea
%Now, using the relations given in Eq. (\ref{relations}) we obtain
yielding,
\ba
\delta \rho^2 = \frac{D(D+1)}{2} \langle \rho \rangle^2 \, .
\ea
Correspondingly, the density contrast is obtained to be
\ba
\label{rho-contrast}
\frac{\delta \rho}{\langle \rho \rangle} =\pm \sqrt{\frac{D(D+1)}{2}} \, ,
\ea
in which the plus sign above correspond to an overdensity while the minus sign represents and underdense region under quantum fluctuations.

The regularized density contrast after setting $D=4-\epsilon$ and setting $\epsilon\rightarrow 0$ is given by
\ba
\label{rho-contrast2}
\Big(\frac{\delta \rho}{\langle \rho \rangle}\Big)_{\mathrm{reg}} =\pm \sqrt{10} \, .
\ea
This agrees exactly with the results in \cite{Firouzjahi:2022xxb} and \cite{Firouzjahi:2022vij} obtained for heavy fields  in flat  as well as in black-hole backgrounds.
However, from the above analysis we see that the result (\ref{rho-contrast2}) is general and is independent of the mass of the field. The fact that the density contrast is independent of the mass and only depends on the dimensionality of the spacetime (as given in Eq. (\ref{rho-contrast})) is an intriguing result. 
As argued in \cite{Firouzjahi:2022xxb}, the fact that $\delta \rho_v \sim 
\langle \rho_v \rangle$ indicates that the distribution of the vacuum zero pint energy is non-linear and non-perturbative, yielding to an inhomogeneous and anistropic background   on small scales, see also \cite{Wang:2017oiy, Cree:2018mcx, Wang:2019mbh, Wang:2019mee} for a similar interpretation.

Having calculated the density contrast, it is also instructive to calculate the contrast in pressure $\frac{\delta P}{\langle P \rangle}$. We have seen that
$\langle P \rangle = - \langle \rho \rangle$ so one may naively expect that the formula (\ref{rho-contrast}) should hold for the pressure contrast as well.  However, there are subtlety here, yielding to a different result. Decomposing the three components of $P= P_1+ P_2 +P_3$ as
\begin{equation}\label{p1-00}
 P_1\equiv \frac{(\partial_{t}\Phi)^2}{2}\,,~~~~~~P_2\equiv \frac{3-D}{2(D-1)} g^{i j} \partial_i \Phi \partial_j \Phi\,,~~~~~~~~P_3\equiv -\frac{m_{\Phi}^2}{2} \Phi^2\,,
\end{equation}
and comparing with Eq. (\ref{rhoi}), we see that
\ba\label{p1-p3}
P_1 = \rho_1 \, ,
\quad \quad
P_2 = \frac{3-D}{D-1} \rho_2 \, ,\quad \quad P_3= -\tilde \rho_3 \, .
\ea
Indeed, the changes in $P_2$  compared to corresponding value for
$\rho_2$  yield to a different result for the pressure contrast.
Following the same steps as above, we obtain
\ba\label{deltaP1-p3}
\delta P^2 &=& \delta P_1^2 +  \delta P_2^2 + \delta P_3^2 \,  \nonumber\\
&=&   2 \big( \langle P_1 \rangle^2 + \langle  P_3 \rangle^2 \big)
+  \frac{2}{D-1}  \langle P_2 \rangle^2 \nonumber\\
&=& \frac{D^3- 5 D +8}{2 (D-1)}  \langle P \rangle^2 \, .
\ea
Correspondingly, the pressure contrast is
\ba
\label{P-contrast}
\frac{\delta P}{\langle P \rangle} =\pm \sqrt{\frac{D^3- 5 D +8}{2 (D-1)}} \, .
\ea

It is instructive to calculate $\frac{\delta P}{\delta \rho}$. Combining
Eqs. (\ref{P-contrast}) and (\ref{rho-contrast}), and noting that $\langle P \rangle = - \langle \rho \rangle$, we obtain
\ba
\label{rho-Pa}
\frac{\delta P}{\delta \rho} = \pm \sqrt{\frac{D^3- 5 D +8}{D(D^2-1)}} \, .
\ea
After setting $D=4-\epsilon$ with $\epsilon \rightarrow 0$, we obtain
\ba
\label{rho-P1}
\Big( \frac{\delta P}{\delta \rho}\Big)_{\mathrm{reg}} = \pm \sqrt{\frac{13}{15}} \, .
\ea

To find a physical interpretation for the meaning of the above ratio, let us treat the
vacuum zero point energy as a cosmic fluid with the equation of state $w\equiv \frac{P}{\rho}$. At the background level, from Eq. (\ref{rho-P})
 we have $w=-1$ as expected from a vacuum zero point energy.  On the other hands, at the perturbation level Eq. (\ref{rho-P1}) suggests that the effective equation of state is   $w=- \sqrt{\frac{13}{15}}$ so the repulsive strength of the dark energy is slightly reduced. This may be interpreted due to quantum nature of the zero point fluctuations unlike the usual picture that the vacuum zero point energy (i.e. cosmological constant) is uniformly distributed in the fabric of spacetime with a uniform equation of state $w=-1$. Having said this, we comment that the above discussion about the effective equation of state of the vacuum zero point energy
 is only qualitative and care must be taken about its cosmological implications.

While the above analysis indicates that the density contrast of the vacuum zero point energy is large, but one should also look at the correlation length $L$ of these perturbations. %This was studied in \cite{Firouzjahi:2022xxb} in which 
The correlation length of zero point fluctuations was studied in \cite{Firouzjahi:2022xxb} in which it is obtained to be at the order $L \sim m^{-1}$. Therefore, for heavy field with $m \gg H$, the correlation length is deep inside the horizon. On the other hand, for a light field with $m \lesssim H$, the correlation length is comparable to or larger than the Hubble radius. The fact that we have large density contrast with long mode perturbations (for light field) can have interesting cosmological implications. %which we would like to investigate elsewhere. 

%%%%%%%%%%%%%%%%%%%%%%%%%%%%%%%%%%%%%%%%%%

\subsection{Measure of Skewness and non-Gaussianity}\label{Skewnessness}

As the distribution of the energy momentum tensor field can be asymmetric and non-Gaussian, it is instructive to calculate the skewness associated to vacuum zero point fluctuations  measured by
$\delta \rho^3 \equiv \langle \rho^3 \rangle - \langle \rho \rangle^3$. A large  value of $\delta \rho^3$ indicates that the system is highly non-Gaussian. Since the source of the energy is quantum fluctuations, it would not be surprising that the system be highly non-Gaussian.

 Following a similar approach as has been employed for calculating the variances,
 one can calculate $\delta \rho^3$. To do this, one can check that (see
 Appendix \ref{contractions} for further details)
\ba
\langle \rho_i^2 \rho_j \rangle=\langle\rho_i^2 \rangle\langle\rho_j \rangle
\quad \quad i\neq j \, ,
\ea
and
\ba
\langle \rho_i \rho_j \rho_k \rangle =  \langle \rho_i  \rangle \langle \rho_j \rangle\langle \rho_k \rangle \quad \quad i\neq j\neq k \, .
\ea
The above relations simplify $\delta \rho^3$ greatly, yielding
\ba
\label{deltarho3-1}
\delta {\rho ^3} = \delta {\rho_1 ^3}+   \delta {\rho_2 ^3}+  \delta {\tilde \rho_3 ^3}
+ 3  \langle \rho_1  \rangle \big( \delta \rho_2^2 + \delta \tilde \rho_3^2 \big)
+ 3  \langle \rho_2  \rangle \big( \delta \rho_1^ 2+ \delta \tilde \rho_3^2 \big)
+ 3  \langle \tilde\rho_3  \rangle \big( \delta \rho_1^2 + \delta  \rho_2^2 \big) \, ,
\ea
 with $\delta \rho_i^2$ given in Eq.  (\ref{var-i}).

Now our job is to calculate $\langle \rho_i^3\rangle$ and then $\delta \rho_i^3$.
Performing various contractions, one can show that the following relations hold (see Appendix \ref{contractions} for further details):
\begin{equation}
\label{deltarho3-ac}
\left\langle {\rho _1^3} \right\rangle  = 15{\left\langle {{\rho _1}} \right\rangle ^3}, \quad \quad \left\langle {\tilde \rho _3^3} \right\rangle  = 15{\left\langle {{\tilde \rho _3}} \right\rangle ^3}\, ,
\end{equation}
while
\ba\label{deltarho3-b}
\left\langle {\rho _2^3} \right\rangle  = {\left\langle {{\rho _2}} \right\rangle ^3}
\Big[1+ \frac{6}{D-1} + \frac{8}{(D-1)^2}
\Big] \, ,
\ea
yielding
 \ba
 \label{var3}
 \delta \rho_1^3 = 14 \langle \rho_1 \rangle^3 \, , \quad \quad
 \delta \rho_2^3 = \Big( \frac{6}{D-1}  + \frac{8}{(D-1)^2} \Big)\langle \rho_2 \rangle^3 \, , \quad \quad
 \delta \tilde\rho_3^3 = 14 \langle \tilde \rho_3 \rangle^3 \, .
 \ea
Plugging the above values of $\delta  \rho_i^3$ in Eq. (\ref{deltarho3-1}), using Eq.  (\ref{var-i}) for $\delta \rho_i^2$ and Eq. (\ref{relations}) expressing
$\langle \rho_i\rangle$ in terms of $\langle \rho\rangle$ we finally obtain
\ba
\frac{\delta \rho^3}{\langle \rho \rangle ^3} = \frac{1}{2}(2D^3+3D^2+D+4) \, .
\ea

Plugging $D=4$ with $\epsilon\rightarrow 0$ yields
\ba
\label{non-G}
\frac{\delta \rho^3}{\langle \rho \rangle ^3}  = 92 \, .
\ea
Therefore, we see that the distribution of the energy density is highly non-Gaussian.

Similarly, for $\delta P^3$ we obtain
\ba
\frac{\delta P^3}{\langle P \rangle ^3}=  \frac{2 D^5 - D^4 - 3 D^3 + D^2 -11 D + 28}{2 (D-1)^2}
\ea
Substituting $D=4-\epsilon$ with $\epsilon\rightarrow 0$ gives
\ba
\label{non-G1}
 \frac{\delta P^3}{\langle P \rangle ^3}  =\frac{800}{9} \, .
\ea
We see that the distribution of $P$ is highly non-Gaussian as well.
%and correspondingly
%\ba
%\label{non-G1}
%\color{blue} \frac{\delta P^3}{\delta \rho ^3}=   -\frac{200}{207} \, .
%\ea

%%%%%%%%%%%%%%%%%%%%%%%%%%%%%%%%%%%%%%%%%%
\section{Summary and Discussions}
\label{summary}

In this work we have studied the vacuum zero point energy of scalar field in dS background. We have allowed the scalar field to have an arbitrary mass parameterized via $\beta = \frac{m}{H}$ and
with the conformal coupling $\xi$. To calculate the vacuum zero point energy and its fluctuations we have employed the dimensional regularization scheme in $D=4-\epsilon$ spacetime. Performing the analysis in a general $D$ dimension, the regularized physical quantities are read off after the subtraction of the divergent
$\frac{1}{\epsilon}$ terms via appropriate counter terms.

We have calculated $\langle \rho_v \rangle, \langle P_v \rangle$ and
$\langle T_v \rangle$. We have shown that
$\langle \rho \rangle_{\mathrm{reg}}=- \langle P \rangle_{\mathrm{reg}}$  as expected from the local Lorentz invariance but the classical relations
$\langle T_v \rangle_{\mathrm{reg}}=- 4\langle P_v \rangle_{\mathrm{reg}}$ is anomalous under quantum corrections. The anomalous correction is given by the factor
${\cal A}$ defined in Eq. (\ref{anomaly}) which vanishes only when $\beta=0$ or
$\xi= \frac{1}{6}- \frac{\beta^2}{12}$.
We have looked at $\langle \rho_v \rangle_{\mathrm{reg}}$ and $\langle T_v \rangle_{\mathrm{reg}}$ in various limits of the parameters space $(\beta, \xi)$. It is shown that for a massless scalar field with $\xi=0$, the vacuum zero point energy is $\langle \rho \rangle_{\mathrm{reg}} \propto H^4 \propto R^2$
which is the hallmark of quantum fields in a curved spacetime. In addition
$\langle T_v \rangle_{\mathrm{reg}} = - 4 \langle \rho_v \rangle_{\mathrm{reg}} \propto H^4 \propto R^2 $ so we have the usual trace anomaly when $m=0$ and $\xi \neq \frac{1}{6}$. On the other hand, the trace anomaly disappears in the conformal massless limit $\beta=0, \xi = \frac{1}{6}$. 

We have shown that for the heavy fields with 
$\beta \gg 1$, the value of $ \langle \rho_v \rangle_{\mathrm{reg}} \sim m^4$
agrees with its value in a flat background plus the subleading $m^2 H^2$ corrections. This is consistent with the physical expectation since the energy density is a local property of the spacetime and the equivalence principle requires that  its value should agree with the corresponding value in a flat background up to subleading corrections.

We have calculated the energy density contrast and the pressure contrast 
for the case when $\xi=0$. In particular, it is shown that
for both massive and massless case $\frac{\delta \rho_v}{\langle \rho_v \rangle} = \pm \sqrt{10}$ as first observed in \cite{Firouzjahi:2022xxb} in the limit of the heavy fields in dS background. This indicates that the distribution of the vacuum zero point energy is non-linear. This can have interesting implications for the cosmological constant problem, indicating a very inhomogeneous and anisotropic background on small scales.
In addition, we have shown that $\frac{\delta P^2}{\delta \rho^2} \neq 1$
indicating a complicated picture for the effective equation of state associated to the vacuum zero point energy as a cosmic fluid.  Since the statistical distribution of the vacuum zero point energy can be asymmetric and non-Gaussian we have calculated $\delta \rho^3$ as a measure of skewness and non-Gaussianity. It is shown that $\delta \rho_v^3 \sim \langle \rho_v \rangle^3$ so the distribution of the vacuum zero point energy is highly non-Gaussian.
%The same conclusion is also obtained for the pressure in which $\delta P_v^3 \sim \langle P_v \rangle^3$.

To simplify the analysis we have worked in the test field limit in which the resultant vacuum zero point energy density is assumed to be much smaller than the background dS energy density. For this picture to hold, we require  $\beta < \sqrt{\frac{M_P}{H}}$. If the field is superheavy, then one can not neglect the backreaction of the vacuum zero point energy on the background geometry.
In addition, to solve the mode function analytically, we have neglected the 
self-coupling of the scalar field such as the $\lambda \Phi^4$ interaction. It would be interesting to extend the above analysis to the case where the field has a 
self-interaction and to see if the above conclusions about the vacuum energy density and its perturbations hold for an interacting field theory as well.

%\newpage
\vspace{0.5cm}
   
 {\bf Acknowledgments:}  We are grateful to   Richard Woodard, Misao Sasaki, Xingang Chen and Mohammad Ali Gorji  for useful  discussions and comments on the draft.  
H. F. is  partially supported by the ``Saramadan" Federation of Iran.

\appendix

\section{Multiple Contractions Using Wick Theorem}
\label{contractions}

In this Appendix we outline the steps which are required to calculate
higher expectations like $\langle \rho_i^2\rangle$ and $\langle \rho_i^3\rangle$
for the variance in section \eqref{Variancedelta2}  and  skewness in section \eqref{Skewnessness}. While the analysis can be done with brute force
in Fourier space but the results can be obtained in real space as well,
using the Wick theorem \cite{Wick:1950ee}. The main ingredient is that $\Phi(x)$ is a Gaussian random field (i.e. a free field) so to calculate higher correlations involving
$\langle \Phi(x) ^n\rangle$ with $n > 3$,
one can use the Wick theorem to reduce them to
combinations of $\langle \Phi(x)^2\rangle$.

Let us start with $ \rho_1= \frac{1}{2} \dot \Phi^2$, yielding
\begin{equation}\label{rho1-app}
\langle \rho _1^2\rangle  = \frac{1}{4}\langle \dot \Phi (x)\dot \Phi (y)\dot \Phi (z)\dot \Phi (w)\rangle\,,
\end{equation}
with the understanding that at the end $x=y=z=w$. Employing the Wick theorem for the Gaussian field $\dot \Phi(x)$, we obtain
\ba
\langle \dot \Phi (x)\dot \Phi (y)\dot \Phi (z)\dot \Phi (w)\rangle = \langle{\dot \Phi (x)\dot \Phi (y) \rangle} \langle{\dot \Phi (z)\dot \Phi (w) \rangle} +
\langle{\dot \Phi (x)\dot \Phi (z) \rangle} \langle{\dot \Phi (y)\dot \Phi (w) \rangle}+ \langle{ \dot \Phi (x)\dot \Phi (w) \rangle} \langle{\dot \Phi (y)\dot \Phi (z) \rangle} .
\nonumber
\ea
Now, setting $x=y=z=w$, from the above three terms we simply obtain
\ba
\langle \rho _1^2\rangle= 3\langle \rho _1\rangle^2\, .
\ea
In a similar manner, we also obtain $\langle \tilde{\rho} _3^2\rangle= 3\langle \tilde \rho _3\rangle^2$.

On the other hand, the analysis for $\langle \rho _2^2\rangle$ is somewhat non-trivial as we have spatial derivatives. More specifically,
\begin{equation}
\label{rho2-app}
\langle \rho _2^2\rangle  =\frac{1}{4}g^{ij}g^{kl}  \langle  {\Phi _{,i}(x)}{\Phi _{,j}}(x){\Phi _{,k}}(y){\Phi _{,l}}(y)\rangle\, .
\end{equation}
Using the Wick theorem, the result can be written as follows:
\ba
\label{rho2-app2}
\langle \rho _2^2\rangle  &=& \frac{1}{4}g^{ij}g^{kl}  \Big[
\langle  {\Phi _{,i}(x)}{\Phi _{,j}}(x)\rangle  \langle  {\Phi _{,k}}(y){\Phi _{,l}}(y)\rangle
+ 2 \langle  {\Phi _{,i}(x)}{\Phi _{,j}}(y)\rangle  \langle  {\Phi_{,k}}(x){\Phi_{,l}}(y)\rangle \Big] \nonumber\\
&=&  \langle \rho_2 \rangle^2 + \frac{1}{2 } \langle  {\Phi _{,i}}(x){\Phi _{,j}}(x) \rangle
\langle  {\Phi ^{,i}}(x){\Phi ^{,j}}(x) \rangle \, .
\ea
Now,  using the isotropy of the background, one can write
\ba
\langle  {\Phi _{,i}}(x){\Phi _{,j}}(x) \rangle= c \delta_{i j} \langle  {(\nabla \Phi})^2  \rangle \, .
\ea
To obtain the coefficient $c$, we contract the above expression with $\delta^{ij}$
and noting that the spatial dimension  is $D-1$, we obtain
\ba
\label{c-val}
c= \frac{1}{D-1} \, ,
\ea
and correspondingly,
\ba
\label{identity}
\langle  {\Phi _{,i}}(x){\Phi _{,j}}(x) \rangle=  \frac{\delta_{i j}}{D-1}  \langle  {(\nabla \Phi})^2  \rangle  = \frac{2}{D-1} \langle \rho_2 \rangle\, .
\ea
Plugging the above result in Eq. (\ref{rho2-app2}) we obtain
\ba
\label{rho2-app3}
\langle \rho _2^2\rangle=   \langle \rho_2 \rangle^2  + \frac{2}{(D-1)}\langle \rho _2\rangle^2 \, .
\ea

Now for the higher orders in a general manner from
the Wick theorem we obtain
\ba
\Big \langle    \Phi ({x_1})\Phi ({x_2}) \ldots \Phi ({x_{2n}}) \Big \rangle
= (2n-1) (2n-3) \langle \Phi(x)^2 \rangle \, .
\ea
In particular, to calculate $\langle \rho _1^3\rangle$ and $\langle \tilde \rho _3^3\rangle$
with $2n=6$, the symmetry factor is $5\times 3=15$ yielding to
Eq. \eqref{deltarho3-ac}.

Now to calculate $\langle \rho _2^3\rangle$, we note that
\ba
\big \langle  \big(\nabla \Phi \big)^6 \big \rangle
= \big \langle  \Phi_{,i}  \Phi^{,i} \,  \Phi_{,j} \Phi^{,j}  \, \Phi_{,k} \Phi^{,k} \rangle  \, .
\ea
There are three types of contractions. The first type is that only identical indices contract to each other. There is only one way for this type of contraction, yielding  simply the result $\langle \rho _2\rangle^3$. The other contraction is that two different indices contract with each other while the remaining two identical indices contract with each other, like $(ii), (jk), (jk)$. There are 6 different ways to do this.
Then using the identity (\ref{identity}) this yields the total contribution
$\frac{6}{D-1} \langle \rho _2\rangle^3$. The last type is to contract all different indices like $(ij), (jk), (ki)$. There are 8 possible ways to do it, and using
the identity (\ref{identity}), this yields the total contribution
$\frac{8}{(D-1)^2} \langle \rho _2\rangle^3$. Combining all, we obtain
\ba
\left\langle {\rho _2^3} \right\rangle  = {\left\langle {{\rho _2}} \right\rangle ^3}
\Big[1+ \frac{6}{D-1} + \frac{8}{(D-1)^2}
\Big] \, ,
\ea
as reported in Eq. (\ref{deltarho3-b}). 

The last thing to show is that for $i\neq j$, $\langle \rho_i \rho_j \rangle=\langle\rho_i \rangle\langle\rho_j \rangle$ and $\langle \rho_i^2 \rho_j \rangle=\langle\rho_i^2 \rangle\langle\rho_j \rangle$ while $\langle \rho_i \rho_j \rho_k \rangle =  \langle \rho_i  \rangle \langle \rho_j \rangle\langle \rho_k \rangle$ for $i \neq j \neq k$. 
To verify these identities, we note that
\ba
\langle \dot \Phi(x) \Phi(x) \rangle= \frac{1}{2}\frac{d}{d t} \langle \Phi(x)^2 \rangle =0 .
\ea
The final equality holds  because, as we have seen in the main text,
$ \langle \Phi(x)^2 \rangle \propto
 \langle  \rho_3 \rangle$ is a constant. This is understandable since the dS spacetime is a maximally symmetric space so $ \langle  \Phi(x)^2 \rangle $ is
independent of $x^\mu$. Similarly, one has
\ba
\langle \partial_i  \Phi(x) \Phi(x) \rangle= \frac{1}{2} \partial_i \langle \Phi(x)^2 \rangle =0 \, .
\ea
Equipped with the above two identities and using the fact that the background is isotropic one can show by direct examinations that for
$i\neq j$, $\langle \rho_i \rho_j \rangle=\langle\rho_i \rangle\langle\rho_j \rangle$ and $\langle \rho_i^2 \rho_j \rangle=\langle\rho_i^2 \rangle\langle\rho_j \rangle$ while for $i \neq j \neq k$, $\langle \rho_i \rho_j \rho_k \rangle =  \langle \rho_i  \rangle \langle \rho_j \rangle\langle \rho_k \rangle$.

{}

\end{document}